\documentclass[aps,prc,amsfonts,preprintnumbers,superscriptaddress,showpacs,nofootinbib]{revtex4}

\usepackage{graphics}
\usepackage{amsmath}
\usepackage{amssymb}
\usepackage{psfrag}
\usepackage{epsfig}
\usepackage{epsf}
\usepackage{float}
\usepackage{slashed}
\allowdisplaybreaks

\newcommand{\fet}[1]{\mbox{\boldmath $#1$}}

\newcommand{\beq}{\begin{equation}}
\newcommand{\eeq}{\end{equation}}
\newcommand{\beqa}{\begin{eqnarray}}
\newcommand{\eeqa}{\end{eqnarray}}
\newcommand{\nn}{\nonumber \\ }

\begin{document}

\title{Box diagram contribution to the axial two-nucleon current}

\author{H.~Krebs}
\email[]{Email: hermann.krebs@rub.de}
\affiliation{Ruhr-Universit\"at Bochum, Fakult\"at f\"ur Physik und
        Astronomie, Institut f\"ur Theoretische Physik II,  D-44780
        Bochum, Germany}
\author{E.~Epelbaum}
\email[]{Email: evgeny.epelbaum@rub.de}
\affiliation{Ruhr-Universit\"at Bochum, Fakult\"at f\"ur Physik und
        Astronomie, Institut f\"ur Theoretische Physik II,  D-44780
        Bochum, Germany}
\author{U.-G.~Mei{\ss}ner}
\email[]{Email: meissner@hiskp.uni-bonn.de}
\affiliation{Helmholtz-Institut~f\"{u}r~Strahlen-~und~Kernphysik~and~Bethe~Center~for~Theoretical~Physics,
~Universit\"{a}t~Bonn,~D-53115~Bonn,~Germany}
\affiliation{Institute~for~Advanced~Simulation,~Institut~f\"{u}r~Kernphysik
and J\"{u}lich~Center~for~Hadron~Physics, Forschungszentrum~J\"{u}lich,~D-52425~J\"{u}lich,~Germany}
\affiliation{Tbilisi State University, 0186 Tbilisi, Georgia}
\date{\today}

\begin{abstract}
Recently, we have worked out the axial two-nucleon current operator to
leading one-loop order in chiral effective field theory using the method of
unitary transformation. Our final expressions, however, differ from the ones
derived by the JLab-Pisa group using time-ordered perturbation
theory~(Phys.\ Rev.\ C {\bf 93}, no. 1, 015501 (2016)
  Erratum: [Phys.\ Rev.\ C {\bf 93}, no. 4, 049902 (2016)]
  Erratum: [Phys.\ Rev.\ C {\bf 95}, no. 5, 059901 (2017)]). 
In this paper we consider the box diagram contribution to the axial
current and demonstrate that the results obtained using the two
methods are unitary equivalent at the Fock-space level. We  
adjust the unitary phases by matching the corresponding two-pion
exchange nucleon-nucleon potentials and rederive the box diagram
contribution to the axial current operator following the approach of
the JLab-Pisa group, thereby reproducing our original result. We provide a
detailed information on the calculation including the relevant
intermediate steps in order to facilitate a clarification of this
disagreement. 
\end{abstract}

\pacs{13.75.Cs,21.30.-x}

\maketitle

\vspace{-0.2cm}

\section{Introduction}
Nuclear axial-vector current operators have been first addressed in 
the framework of chiral effective field theory (EFT) in Ref.~\cite{Park:1993jf}. The dominant
single-nucleon contribution emerges at order $Q^{-3}$, with $Q$  denoting
the expansion parameter of chiral EFT, from the standard
Gamow-Teller operator. Contributions to the exchange axial charge and
current operators for general kinematical conditions have been
recently worked out to order $Q$ by the
Bochum-Bonn group \cite{Krebs:2016rqz}  using the method of unitary
transformation and independently by the JLab-Pisa group \cite{Baroni:2015uza} using
time-ordered perturbation theory. The latter approach relies on the
transfer matrix and defines the effective potential and current
operators by subtracting the corresponding iterative contributions.
Generally, nuclear interactions derived using different methods
are expected to be equivalent modulo off-shell effects.
However, a direct comparison of the results for e.g.~the two-nucleon
two-pion-exchange contributions proportional to $g_A^5$,
where $g_A$ denotes the nucleon
axial coupling constant, given in~Eqs.~(5.29) and (5.31) of our work \cite{Krebs:2016rqz}
and those in Eqs.~(7.4), (7.5) of Ref.~\cite{Baroni:2015uza}, see also
Eqs. (16), (17) of Ref.~\cite{Baroni:2016xll}, lets one conclude that both results
cannot be unitarily equivalent for the class of unitary
transformations considered in~\cite{Krebs:2016rqz}, thus indicating
that at least one of the calculations should be incorrect (unless we
have misinterpreted the approach and/or conventions of Refs.~\cite{Baroni:2015uza,Baroni:2016xll}). 
To shed light on this issue  and to enable a more direct comparison
between the two approaches, we
present in this paper a detailed calculation of the contribution of 
the box diagram e8 in Fig.~4 of~\cite{Baroni:2015uza}
to the exchange axial current density, which leads to the already
mentioned problematic terms
$\propto g_A^5$.
Specifically, we re-derive the corresponding
expressions using the  method of the JLab-Pisa group, thereby
reproducing our original results. To facilitate the error diagnostics and
a more detailed comparison, we also provide various intermediate-stage expressions of our
calculations. 

Our paper is organized as follows.  Given that 
nuclear potentials and currents are scheme-dependent quantities, we
first need to clarify the relation between the interactions obtained
using both methods. To this aim, we focus in section~\ref{sec2} on the
case without external sources and employ the method of the JLab-Pisa
group to derive the expressions for the two-pion exchange
potential, which turn out to coincide in both approaches. This allows us to
unambiguously fix the phases of two unitary transformations on the purely nucleonic
subspace of the Fock space that appear at this chiral order. We
then give the Fock-space expression for the effective potential of the JLab-Pisa
group up to order $Q^3$ in terms of the pion-nucleon vertex
$\propto g_A$ and the corresponding energy denominators, including
relativistic corrections that are not related to the expansion of the
$g_A$-vertex.
Having fixed the unitary phases as
described above, we use the JLab-Pisa method to derive the Fock-space
expressions for the axial-vector operators involving  $1$, $3$ and
$5$ pion-nucleon vertices $\propto g_A$.
The resulting expressions are verified to be unitarily equivalent to
the ones obtained in Ref.~\cite{Krebs:2016rqz} using the method of unitary
transformation. Next, in section~\ref{sec3}, we use the derived
Fock-space operators to calculate the non-pion-pole contributions of the box and
crossed-box diagrams to the axial-vector
current. We give expressions for the current before evaluating the
loop integrals, perform the Passarino-Veltman reduction of the
relevant tensor integrals and provide explicit expressions for the remaining
scalar integrals. The results of this paper are briefly summarized in
section~\ref{sec4}.

\section{Nuclear forces up to NLO: Fixing the unitary ambiguity }
\label{sec2}

To derive the effective nuclear potential 
we start with Eqs.~(12)-(15) 
of Ref.~\cite{Pastore:2011ip}. The half-off-shell transfer matrix
$T^{(n)}$ and the free Green's function $G_0$ in
these equations depend on the energy $E_i$ of the initial state (see Eq.~(8)
of~\cite{Pastore:2011ip}). Since we focus here on the box diagrams, we
only retain the leading pion-nucleon vertex $\propto g_A$,
and we also do not consider the relativistic corrections to this
vertex. The inversion of the half-off-shell transfer
matrix via Eqs.~(12)-(15) of~\cite{Pastore:2011ip} leads to the
effective potential, that is identical to the one obtained using the so-called
folded-diagram technique \cite{Kuo:1971uze}, see also Ref.~\cite{Krebs:2004st}:
\beqa
v_{\rm FD}^{(0)}&=&-\eta V \frac{\lambda^1}{\omega} V \eta,\nn
v_{\rm FD}^{(1)}&=&\eta V {\cal E} \frac{\lambda^1}{\omega^2} V \eta
-\eta V \frac{\lambda^1}{\omega^2} V \eta H_0 
\eta~,\nn
v_{\rm FD}^{(2)}&=&2\eta V {\cal E} \frac{\lambda^1}{\omega^3} V \eta 
H_0 \eta-\eta V \frac{\lambda^1}{\omega^3} V 
\eta H_0^2 \eta
-\eta V {\cal E}^2 \frac{\lambda^1}{\omega^3} 
V \eta
+\eta V \frac{\lambda^1}{\omega^2} 
V \eta V \frac{\lambda^1}{\omega} 
V \eta
-\eta V \frac{\lambda^1}{\omega} 
V \frac{\lambda^2}{\omega} V 
\frac{\lambda^1}{\omega} V \eta~,\nn
v_{\rm FD}^{(3)}&=&\eta V \frac{\lambda^1}{\omega^3} V \eta H_0 
\eta V \frac{\lambda^1}{\omega} V \eta
+\eta V \frac{\lambda^1}{\omega^3} V \eta 
V \frac{\lambda^1}{\omega} V \eta H_0 
\eta
+\eta V \frac{\lambda^1}{\omega^2} V \eta 
V \frac{\lambda^1}{\omega^2} V \eta H_0 \eta
-\eta V \frac{\lambda^1}{\omega^2} V 
\frac{\lambda^2}{\omega} V 
\frac{\lambda^1}{\omega} V \eta H_0 \eta\nn
&-&\eta V \frac{\lambda^1}{\omega} V 
\frac{\lambda^2}{\omega^2} V 
\frac{\lambda^1}{\omega} V \eta H_0 \eta
-\eta V \frac{\lambda^1}{\omega} V 
\frac{\lambda^2}{\omega} V 
\frac{\lambda^1}{\omega^2} V \eta H_0 \eta
-2\eta V {\cal E} \frac{\lambda^1}{\omega^3} V \eta 
V \frac{\lambda^1}{\omega} V \eta
+\eta V {\cal E} \frac{\lambda^1}{\omega^2} V 
\frac{\lambda^2}{\omega} V 
\frac{\lambda^1}{\omega} V \eta\nn
&-&\eta V \frac{\lambda^1}{\omega^2} V \eta V {\cal E} 
\frac{\lambda^1}{\omega^2} V \eta
+\eta V \frac{\lambda^1}{\omega} V {\cal E} 
\frac{\lambda^2}{\omega^2} V 
\frac{\lambda^1}{\omega} V \eta
+\eta V \frac{\lambda^1}{\omega} V \frac{\lambda^2}{\omega} 
V {\cal E} \frac{\lambda^1}{\omega^2} V \eta~.\label{folded_diagram_veff}
\eeqa
Here and in what follows, $\eta$ ($\lambda^i$) denotes the projection
operator onto the purely nucleonic states of the Fock space (states
involving $i$ pions), $V$ is the operator corresponding to the pion-nucleon vertex $\propto
g_A$\footnote{In our earlier paper \cite{Krebs:2016rqz}, we used for this
  vertex the notation $H_{2,1}^{(1)}$ in order to signify that it
  involves two nucleon fields and one pion field and has the dimension
  $\kappa =1$ as defined in Ref.~\cite{Krebs:2016rqz}. Since we only consider this type of
  vertex here, we choose to employ a simpler notation.}, while
$\omega$ and ${\cal E}$ denote the sum of $n$ pion energies $\omega_i = \sqrt{\vec
  p_i^{\, 2} + M_\pi^2}$ and  the kinetic energy of nucleons in an
intermediate state $\lambda^n$, respectively. 
Further, the superscript ${(n)}$ gives the order $Q$ of the
chiral expansion. Following the approach of Refs.~\cite{Baroni:2015uza,Baroni:2016xll,Pastore:2011ip}, we count
the nucleon mass as $m_N \sim \Lambda_\chi$, with $\Lambda_\chi$
denoting the breakdown scale of chiral EFT. The calculations by the
Bochum-Bonn group employ the counting scheme with $Q/m_N\sim(Q/\Lambda_\chi)^2$.

The manifestly non-hermitean effective potential in Eq.~(\ref{folded_diagram_veff}) is uniquely
determined for a given half-off-shell T-matrix.
Changing the off-shell behaviour of the T-matrix by  adding
terms proportional to $[H_0, X]$, where $X$ is an arbitrary operator,
and applying the same inversion procedure leads to a different
effective potential.
Off-shell changes of the T-matrix can be understood in terms of
similarity transformations of the effective Hamiltonian.
Since the authors of Ref.~\cite{Pastore:2011ip} do
not specify the operator $X$, we need to extract the off-shell
behavior of the T-matrix
from their final expressions for the 
effective potential. To derive nuclear forces in the convention of the
JLab-Pisa group, we apply a series of similarity
transformations on the potential $v_{\rm FD}$ of
Eq.~(\ref{folded_diagram_veff}). We first bring
$v_{\rm FD}$ into a hermitean form by applying a similarity transformation~\cite{Suzuki:PTP1983,Krebs:2004st}
\beqa
v_{\rm Okubo}&=&\big(1+A A^\dagger + A^\dagger A\big)^{1/2} v_{\rm FD}
\big(1+A A^\dagger + A^\dagger A\big)^{-1/2}~.  
\eeqa
Here, $v_{\rm Okubo}$ is precisely the
hermitean potential that is obtained from the underlying pion-nucleon
Hamiltonian $H$ via the unitary transformation introduced by Okubo
\cite{Okubo:1954zz}, 
\beqa
v_{\rm Okubo} =  U^\dagger H U -
H_0,\label{UT_effective_potential_okubo}, \quad \quad
U=\left( \begin{array}{cc} \big(\eta+A^\dagger A\big)^{-1/2} & -
                                                               A^\dagger\big(1+
                                                               A
                                                               A^\dagger\big)^{-1/2}\\
           A\big(1+A^\dagger A)^{-1/2} &
                                         \big(\lambda+A
                                         A^\dagger\big)^{-1/2}
                                         \end{array} \right)\,,
\eeqa
with the operator $A = \lambda A \eta$ satisfying the nonlinear decoupling equation
\beqa
\lambda\big(H - \big[A, H\big] - A H A\big)\eta &=&0~.\label{decoupling_eq_for_A_op}
\eeqa
Notice that the Okubo unitary transformation leads to 
non-factorizable and non-renormalizable nuclear potentials
\cite{Epelbaum:2007us}. For the class of contributions considered in
this work, renormalizability of the nuclear potentials can be restored
by performing additional $\eta$-space unitary transformations 
\beqa
v&=&U_{\rm 12}^\dagger  (v_{\rm Okubo} + H_0 )U_{\rm 12} - H_0~,
\eeqa
with
\beqa
U_{\rm 12}&=&\exp\big(\alpha_1 S_1 + \alpha_2 S_2\big)~,
\eeqa
where the antihermitean operators $S_1$ and $S_2$ are defined in Eq.~(3.25)
of~\cite{Epelbaum:2007us}. To fix the unitary phases $\alpha_1$ and
$\alpha_2$  we match the expression for the two-pion exchange
two-nucleon potential obtained from $v$ with Eq.~(19) of Ref.~\cite{Pastore:2011ip}.
We reproduce the expression in Eq.~(19) of Ref.~\cite{Pastore:2011ip}
provided the  unitary phases $\alpha_1$ and $\alpha_2$ are chosen to be\footnote{Note that there is
a misprint in Eq.~(3.31) of~\cite{Epelbaum:2007us}, a factor of $2$
in front of $\alpha_2$ is missing. The corrected equation reads $\alpha_1=-2\alpha_2=-\frac{1}{2}$.} 
\beqa
\alpha_1&=&-\frac{1}{2}, \quad \alpha_2\,=\,\frac{1}{4}.\label{additional_a1a2}
\eeqa
This particular choice leads to renormalizable
nuclear potentials \cite{Epelbaum:2007us} and is employed also by the Bochum-Bonn group.

The leading relativistic corrections to the nuclear forces are well
known to depend on two arbitrary phases $\bar \beta_8$, $\bar
\beta_9$, see Eq.~(1.4) of Ref.~\cite{Krebs:2019aka} for the definition. In
Ref.~\cite{Friar:1999sj}, the same off-shell ambiguity is expressed in terms of the
phases $\mu$, $\nu$.  To be consistent with the choice made by the
JLab-Pisa group for the one-pion exchange potential, see Eq.~(19) of Ref.~\cite{Pastore:2011ip},
we set $\nu=0$ which corresponds to $\beta_8=\nu/2=0$
in our notation. The second phase needs not be discussed here since we
do not consider relativistic corrections to the $g_A$-vertex.  

With the above choices, we arrive at the Fock-space expressions for the effective
potential that correspond to the convention of the JLab-Pisa group: 
\beqa
v^{(0)}&=& -\eta V \frac{\lambda^1}{\omega} V \eta~,\label{veff_Q0}\\
v^{(1)}&=& \frac{1}{2}\eta V {\cal E} \frac{\lambda^1}{\omega^2} 
V \eta-\frac{1}{2}\eta H_0 \eta V 
\frac{\lambda^1}{\omega^2} V \eta+ \; {\rm h.c.}\;,\label{veff_Q1}\\
v^{(2)}&=&\eta H_0 \eta V {\cal E} \frac{\lambda^1}{\omega^3} 
V \eta
-\frac{1}{2}\eta H_0^2 \eta V \frac{\lambda^1}{\omega^3}
V \eta
-\frac{1}{2}\eta V {\cal E}^2 \frac{\lambda^1}{\omega^3}
V \eta
+\frac{1}{2}\eta V \frac{\lambda^1}{\omega^2} 
V \eta V \frac{\lambda^1}{\omega} 
V \eta
-\frac{1}{2}\eta V 
\frac{\lambda^1}{\omega} V 
\frac{\lambda^2}{\omega} V 
\frac{\lambda^1}{\omega} V \eta \nn
&+& \;{\rm h.c.}\;,\label{veff_Q2}\\
v^{(3)}&=& \alpha_1\bigg(\eta H_0 \eta V \frac{\lambda^1}{\omega} 
V \eta V \frac{\lambda^1}{\omega^3} 
V \eta
-\eta H_0 \eta V 
\frac{\lambda^1}{\omega^3} V \eta V 
\frac{\lambda^1}{\omega} V \eta\bigg)\nn
&+&\alpha_2\bigg(\eta H_0 \eta 
V \frac{\lambda^1}{\omega} V 
\frac{\lambda^2}{\omega} V 
\frac{\lambda^1}{\omega^2} V \eta
-\eta H_0 \eta V \frac{\lambda^1}{\omega^2} V 
\frac{\lambda^2}{\omega} V 
\frac{\lambda^1}{\omega} V \eta\bigg)\nn
&+&\frac{3}{8}\eta H_0 \eta 
V \frac{\lambda^1}{\omega^2} V \eta 
V \frac{\lambda^1}{\omega^2} V \eta
-\frac{1}{2}\eta H_0 \eta V 
\frac{\lambda^1}{\omega^2} V 
\frac{\lambda^2}{\omega} V 
\frac{\lambda^1}{\omega} V \eta
+\frac{1}{2}\eta H_0 \eta V \frac{\lambda^1}{\omega} V \eta 
V \frac{\lambda^1}{\omega^3} V \eta
-\frac{1}{2}\eta H_0 \eta V \frac{\lambda^1}{\omega} 
V \frac{\lambda^2}{\omega^2} V 
\frac{\lambda^1}{\omega} V \eta\nn
&-&\frac{1}{2}\eta H_0 \eta V \frac{\lambda^1}{\omega} V 
\frac{\lambda^2}{\omega} V 
\frac{\lambda^1}{\omega^2} V \eta
+\frac{1}{2}\eta V \frac{\lambda^1}{\omega^3} V \eta H_0 \eta 
V \frac{\lambda^1}{\omega} V \eta
+\frac{1}{8}\eta V \frac{\lambda^1}{\omega^2} V \eta H_0 
\eta V \frac{\lambda^1}{\omega^2} V \eta
-\eta V {\cal E} \frac{\lambda^1}{\omega^3} V \eta 
V \frac{\lambda^1}{\omega} V \eta\nn
&-&\frac{1}{2}\eta V {\cal E} \frac{\lambda^1}{\omega^2} V \eta 
V \frac{\lambda^1}{\omega^2} V \eta
+\eta V {\cal E} \frac{\lambda^1}{\omega^2} V 
\frac{\lambda^2}{\omega} V 
\frac{\lambda^1}{\omega} V \eta
+\frac{1}{2}\eta V \frac{\lambda^1}{\omega} V {\cal E} 
\frac{\lambda^2}{\omega^2} V 
\frac{\lambda^1}{\omega} V \eta+ \; {\rm h.c.}\;.\label{veff_Q3}
\eeqa
We have retained here the dependence on the phases $\alpha_{1,2}$
which should be chosen according to Eq.~(\ref{additional_a1a2}). A comparison
of Eqs.~(\ref{veff_Q0})-(\ref{veff_Q3}) with Eq.~(3.13)
of~\cite{Epelbaum:2005pn}\footnote{In Eq.~(3.13)
  of~\cite{Epelbaum:2005pn}, the author discusses the Yukawa model and not
  chiral effective field theory. This is, however, perfectly
  sufficient for our purpose since
  we are only interested in the box diagram contributions. The only vertex
which is relevant for the current discussion of effective potentials
is the leading one-pion-nucleon vertex $V$ which can be interpreted as
a Yukawa-type interaction.}, obtained using the Okubo transformation, shows
that both results indeed coincide for the case of
$\alpha_1=\alpha_2=0$ as already pointed out above.
We further emphasize that we have neglected all
relativistic corrections  in Eqs.~(\ref{veff_Q1}), (\ref{veff_Q2}) and
(\ref{veff_Q3})
which scale as $1/m_N^n$
with $n>2$ ($n>1$) for operators involving two (four) insertions of
the pion-nucleon vertex $V$. The neglected effects are
of a higher order in the chiral expansion.

We now turn to the axial-vector currents.  
As in the case of the nuclear forces, effective current operators can
be derived by inverting the contribution to the T-matrix $T_5$ that depends
linearly on the corresponding external sources. This in turn is the
approach followed by the JLab-Pisa group to derive the axial-vector current
operator $v_5$, see Eqs.~(3.8)-(3.12) of~\cite{Baroni:2015uza}. 
In these equations the
operators $T_5$ and $G_0$ depend on the initial- and final-state
energies $E_i$ and $E_f$. The relation between the
axial-vector current $v_5(E_f-E_i)$ and the T-matrix $T_5$
is derived in Appendix~\ref{app:transfer_matrix} and given by
\beqa
T_5(E_f, E_i\,)&=& \big(1-v 
G_0(E_f)\big)^{-1} \; v_5(E_f-E_i) \;
\big(1- G_0(E_i) v\big)^{-1}~.\label{v5_FD_inversion}
\eeqa
Here, we have already exploited the fact that the axial-vector
source appears in first order perturbation theory.
The explicit dependence of the interaction on the axial-vector source is hidden in $v_5(E_f-E_i)$.
Its dependence on the energy difference $E_f-E_i$ emerges due to the explicit time-dependence of axial-vector
source.

We now invert Eq.~(\ref{v5_FD_inversion}) in the same
way as done by the JLab-Pisa group\footnote{The explicit energy
  dependence as given in Eq.~(\ref{v5_FD_inversion}) is not spelled out in
  Eqs.~(3.8)-(3.12) of~\cite{Baroni:2015uza}.} to obtain the
axial-vector current operator  using
Eqs.~(\ref{veff_Q0}),  (\ref{veff_Q1}), (\ref{veff_Q2}) and (\ref{veff_Q3}) for the
strong-interaction potential $v$. The resulting somewhat lengthy
Fock-space expressions 
are given in Appendix~\ref{app:axial_vector_fock_space}. Notice that
exactly the same expressions are obtained using the method of unitary
transformation in Ref.~\cite{Krebs:2016rqz}, provided all additional unitary
transformations that depend explicitly on the external axial-vector
sources are switched off, i.e.~$\alpha_i^{ax} = 0$ for all $i$.  This
demonstrates that the current operators derived by the two
groups should be unitarily equivalent. 

\section{Calculation of the box diagram}
\label{sec3}

\begin{figure}[tb]
\vskip 1 true cm
\includegraphics[width=17cm,keepaspectratio,angle=0,clip]{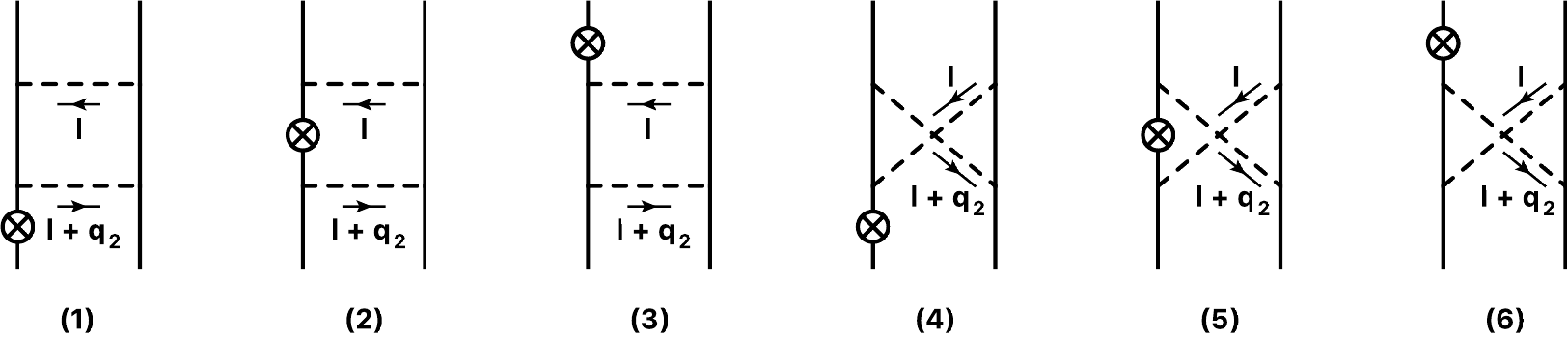}
    \caption{Box diagrams proportional to $g_A^5$. Diagrams obtained 
      by permutation of the nucleon labels $1\leftrightarrow 2$ are not
      shown. Crossed circles denote the coupling of axial-vector
      source $\propto g_A$. The momenta $\vec l$ and $\vec l+ \vec
      q_2$ in the static pion propagators are
      explicitly spelled out. Initial (final) momenta momenta of the nucleons are
      denoted by $\vec p_1$ and $\vec p_2$ ($\vec p_1^{\, \prime}$ and
      $\vec p_2^{\, \prime}$).
      The momentum transfer of the nucleon $j$ is
      defined by $\vec q_j=\vec p_j^{\, \prime}- \vec p_J$ with $j=1,2$.
\label{fig:boxdiagr} 
 }
\end{figure}

We now use the expressions for $v_5$ given in
Appendix~\ref{app:axial_vector_fock_space} to calculate the
contributions of the box diagrams to the axial-vector current operator, which are
visualized in Fig.~\ref{fig:boxdiagr}. While we focus here exclusively on the
non-pion-pole contributions, the Fock-space expressions for the
pion-pole terms are also provided in
Appendix~\ref{app:axial_vector_fock_space}. 
Each diagram shown in Fig.~\ref{fig:boxdiagr} gives rise to a series
of time-ordered graphs, whose contributions can be obtained by
calculating the two-nucleon matrix elements of the operators in
Eq.~(\ref{NP}). Time-ordered graphs associated with any of
the six diagrams in Fig.~\ref{fig:boxdiagr} feature the same sequence of
vertices and thus have the same spin-isospin-momentum structure which
will be given below. We begin with collecting together the energy
denominators for diagrams (1)-(6) and obtain the following result: 
\beqa
{\rm box}\; (1)&=&\frac{2}{\omega_l 
\omega_{\vec{l}+\vec{q}_2}^3}+
\bigg(\alpha_1+\frac{1}{2}\bigg) \left[\frac{4}{\omega_l^3 
\omega_{\vec{l}+\vec{q}_2}}-\frac{4}{\omega_l \omega_{\vec{l}+\vec{q}_2}^3}
\right]\nn
&+&\bigg(\alpha_2-\frac{1}{4}\bigg) \left[\frac{2}{\omega_l^2 
\omega_{\vec{l}+\vec{q}_2} (\omega_{\vec{l}+\vec{q}_2}+\omega_l)}-\frac{2}{\omega_l 
\omega_{\vec{l}+\vec{q}_2}^2 (\omega_{\vec{l}+\vec{q}_2}+\omega_l)}\right]~,\nn
{\rm box}\; (2)&=&-\frac{2}{\omega_l \omega_{\vec{l}+\vec{q}_2}^3}-\frac{2}{\omega_l^3 
\omega_{\vec{l}+\vec{q}_2}}~,\nn
{\rm box}\; (3)&=&\frac{2}{\omega_l^3 
\omega_{\vec{l}+\vec{q}_2}}+
\bigg(\alpha_1+\frac{1}{2}\bigg) \left[\frac{4}{\omega_l 
\omega_{\vec{l}+\vec{q}_2}^3}-\frac{4}{\omega_l^3 \omega_{\vec{l}+\vec{q}_2}}
\right]\nn
&+&\bigg(\alpha_2-\frac{1}{4}\bigg) \left[\frac{2}{\omega_l 
\omega_{\vec{l}+\vec{q}_2}^2 (\omega_{\vec{l}+\vec{q}_2}+\omega_l)}-\frac{2}{\omega_l^2 
\omega_{\vec{l}+\vec{q}_2} (\omega_{\vec{l}+\vec{q}_2}+\omega_l)}\right]~,\nn
{\rm box}\; (4)&=&-\frac{2}{\omega_l^3 \omega_{\vec{l}+\vec{q}_2}}~, \nn
{\rm box}\; (5)&=&\frac{2}{\omega_l \omega_{\vec{l}+\vec{q}_2}^3}+\frac{2}{\omega_l^3 
\omega_{\vec{l}+\vec{q}_2}}~,\nn
{\rm box}\; (6)&=&-\frac{2}{\omega_l \omega_{\vec{l}+\vec{q}_2}^3}~.\label{energy_denominators}
\eeqa
For the choice of the unitary phases $\alpha_1$ and $\alpha_2$ made by our
and by the JLab-Pisa group given in Eq.~(\ref{additional_a1a2}),  all
terms in the square brackets vanish leading to the
factorized results for every diagrams.

Combining the energy denominators  in Eq.~(\ref{energy_denominators})
with the corresponding spin-isospin-momentum structures, the complete
contribution of the box diagrams to the axial-vector current can be
written in the form
\beqa
\vec{\cal A}_{\rm box}^a&=&\sum_{i=1}^6
\int \frac{d^3 l}{(2\pi)^3} \; 
\vec{\cal A}_{{\rm box} \, (i)}^a \; +\; 
(1\,\leftrightarrow \,2)~,
\eeqa
where $a$ is an isospin index and the $\vec{\cal A}_{{\rm box} \, (i)}^a$ are given by 
\beqa
\vec{\cal A}_{{\rm box} \, (1)}^a&=&
\frac{g_A^5 }{64 F_\pi^4 
\omega_l^2 \omega_{\vec{l}+\vec{q}_2}^4}\big(2 i\, [\tau_1\times\tau_2]^a-3 \tau_1^a+2 \tau_2^a
\big) \bigg(\vec{l} \,\big[-i\, \vec{q}_2\cdot \vec{\sigma}_1 \vec{l}\cdot 
\vec{q}_2\times \vec{\sigma}_2-(l^2+ \vec{l}\cdot \vec{q}_2)\vec{q}_2\cdot 
\vec{\sigma}_1\nn
&+&q_2^2 \vec{l}\cdot 
\vec{\sigma}_2-\vec{l}\cdot \vec{q}_2 \vec{q}_2\cdot \vec{\sigma}_2\big]+
\vec{q}_2 \,\big[i \,\vec{l}\cdot \vec{\sigma}_1 \vec{l}\cdot \vec{q}_2\times 
\vec{\sigma}_2+\vec{l}\cdot \vec{q}_2 (\vec{l}\cdot 
\vec{\sigma}_1-\vec{l}\cdot \vec{\sigma}_2)
+l^2 
(\vec{l}\cdot \vec{\sigma}_1+\vec{q}_2\cdot \vec{\sigma}_2)\big]\nn
&+&\vec{\sigma}_1 \,(l^2+\vec{l}
\cdot \vec{q}_2\,) (i \,\vec{l}\cdot 
\vec{q}_2\times \vec{\sigma}_2+l^2+\vec{l}\cdot \vec{q}_2
\,)+\vec{\sigma}_2\big[(\vec{l}\cdot \vec{q}_2)^2-l^2 
q_2^2\big] +i\, \vec{l}\times\vec{q}_2\, (l^2+\vec{l}\cdot 
\vec{q}_2\,) \bigg)~,\\
\vec{\cal A}_{{\rm box} \, (2)}^a&=&
\frac{g_A^5 \left(\omega_{\vec{l}+\vec{q}_2}^2+\omega_l^2
\right) }{64 F_\pi^4 \omega_l^4 \omega_{\vec{l}+\vec{q}_2}^4}
\big(2 \tau_2^a-\tau_1^a\big) \bigg(-\vec{l} \,\big[-i\, \vec{l}\cdot 
\vec{\sigma}_1 \vec{l}\cdot \vec{q}_2\times\vec{\sigma}_2-i \,
\vec{q}_2\cdot \vec{\sigma}_1
\vec{l}\cdot\vec{q}_2\times\vec{\sigma}_2
-\vec{l}\cdot \vec{q}_2 (2 \,\vec{l}\cdot
\vec{\sigma}_1\nn
&+&\
\vec{q}_2\cdot \vec{\sigma}_1+\vec{q}_2\cdot 
\vec{\sigma}_2)-l^2 (2 \vec{l}\cdot 
\vec{\sigma}_1+\vec{q}_2\cdot \vec{\sigma}_1)+q_2^2
 \vec{l}\cdot \vec{\sigma}_2\big]-i\, \vec{l}\times\vec{q}_2 \big[-(\vec{l}\cdot 
\vec{\sigma}_1) (\vec{l}\cdot \vec{\sigma}_2+\vec{q}_2\cdot 
\vec{\sigma}_2)\nn
&+&(l^2+\vec{l}\cdot \vec{q}_2) (\vec{\sigma}_1\cdot 
\vec{\sigma}_2+1)\big] -i\, \vec{l}\times\vec{\sigma}_2\, \big[q_2^2
\vec{l}\cdot \vec{\sigma}_1+\vec{l}\cdot \vec{q}_2 (\vec{l}\cdot 
\vec{\sigma}_1-\vec{q}_2\cdot \vec{\sigma}_1)-l^2 
\vec{q}_2\cdot \vec{\sigma}_1\big] \nn
&-&\vec{q}_2 
\big[l^2 (\vec{q}_2\cdot \vec{\sigma}_2-\vec{l}\cdot 
\vec{\sigma}_1)-\vec{l}\cdot \vec{q}_2 (\vec{l}\cdot 
\vec{\sigma}_1+\vec{l}\cdot \vec{\sigma}_2)\big]-\vec{\sigma}_1 (l^2+\vec{l}\cdot \vec{q}_2\,)^2 -\vec{\sigma}_2 \big[(\vec{l}\cdot 
\vec{q}_2)^2-l^2 q_2^2\big] 
\bigg)~,\\
\vec{\cal A}_{{\rm box} \, (3)}^a&=&
\frac{g_A^5 }{64 F_\pi^4 \omega_l^4 \omega_{\vec{l}+\vec{q}_2}^2}\big(2 i [\tau_1\times\tau_2]^a+3 \tau_1^a-2
\tau_2^a
\big) \bigg(-\vec{l} \,\big[i\, (\vec{l}\cdot \vec{\sigma}_1+\vec{q}_2\cdot 
\vec{\sigma}_1) \vec{l}\cdot \vec{q}_2\times 
\vec{\sigma}_2+(l^2+\vec{l}\cdot \vec{q}_2) 
\vec{q}_2\cdot \vec{\sigma}_1\nn
&+& q_2^2 \vec{l}\cdot 
\vec{\sigma}_2-\vec{l}\cdot \vec{q}_2 \vec{q}_2\cdot 
\vec{\sigma}_2\big]-i \, \vec{l}
\times\vec{q}_2\,\big[-\vec{l}\cdot \vec{\sigma}_1 (\vec{l}\cdot 
\vec{\sigma}_2+\vec{q}_2\cdot \vec{\sigma}_2)+(l^2+\vec{l}\cdot 
\vec{q}_2) (\vec{\sigma}_1\cdot 
\vec{\sigma}_2 + 1)\big] \nn
&-&i\, \vec{l}
\times\vec{\sigma}_2\, \big[(\vec{l}\cdot \vec{q}_2+q_2^2) 
\vec{l}\cdot \vec{\sigma}_1-(l^2+\vec{l}\cdot \vec{q}_2)
 \vec{q}_2\cdot \vec{\sigma}_1\big] -\vec{q}_2 \big[l^2 (\vec{q}_2\cdot 
\vec{\sigma}_2-\vec{l}\cdot \vec{\sigma}_1)-\vec{l}\cdot \vec{q}_2 
(\vec{l}\cdot \vec{\sigma}_1+\vec{l}\cdot 
\vec{\sigma}_2)\big]\nn
&-&\vec{\sigma}_1(l^2+\vec{l}\cdot \vec{q}_2)^2 
-\vec{\sigma}_2\,\big[(\vec{l}\cdot \vec{q}_2)^2-l^2 
q_2^2\big] \bigg)~,\\
\vec{\cal A}_{{\rm box} \, (4)}^a&=&
\frac{g_A^5 }{64 F_\pi^4 
\omega_l^4 \omega_{\vec{l}+\vec{q}_2}^2}\big(2 i \,[\tau_1\times\tau_2]^a +3 \tau_1^a+2 \tau_2^a
\big) \bigg(\vec{l} \,\big[i\, \vec{q}_2\cdot \vec{\sigma}_1 \vec{l}\cdot 
\vec{q}_2\times \vec{\sigma}_2+(l^2+\vec{l}\cdot \vec{q}_2) \vec{q}_2\cdot 
\vec{\sigma}_1\nn
&-&q_2^2 \vec{l}\cdot 
\vec{\sigma}_2+\vec{l}\cdot \vec{q}_2 \vec{q}_2\cdot 
\vec{\sigma}_2\big]+\vec{q}_2 \big[-i \,\vec{l}\cdot \vec{\sigma}_1 
\vec{l}\cdot \vec{q}_2\times \vec{\sigma}_2+\vec{l}\cdot \vec{q}_2 
(\vec{l}\cdot \vec{\sigma}_2-\vec{l}\cdot 
\vec{\sigma}_1)-l^2 (\vec{l}\cdot 
\vec{\sigma}_1+\vec{q}_2\cdot \vec{\sigma}_2)\big]\nn
&+&\vec{\sigma}_1 \,(l^2+\vec{l}\cdot 
\vec{q}_2) (i\, \vec{l}\cdot 
\vec{q}_2\times \vec{\sigma}_2+\vec{l}\cdot \vec{q}_2+l^2 
)+\vec{\sigma}_2\,\big[l^2 q_2^2-(\vec{l}
\cdot \vec{q}_2)^2\big] -i\, \vec{l}\times\vec{q}_2\, \big[\vec{l}\cdot 
\vec{q}_2+l^2\big] \bigg),\\
\vec{\cal A}_{{\rm box} \, (5)}^a&=&
\frac{g_A^5 \left(\omega_{\vec{l}+\vec{q}_2}^2+\omega_l^2
\right) }{64 F_\pi^4 \omega_l^4 \omega_{\vec{l}+\vec{q}_2}^4}
\big(2 \tau_2^a+\tau_1^a\big) \bigg(\vec{l}\, \big[i \,\vec{l}\cdot 
\vec{\sigma}_1 \vec{l}\cdot \vec{q}_2\times 
\vec{\sigma}_2+\vec{l}\cdot \vec{q}_2 (2 \vec{l}\cdot \vec{\sigma}_1+
\vec{q}_2\cdot \vec{\sigma}_1-\vec{q}_2\cdot 
\vec{\sigma}_2)\nn
&+&l^2\, (2 \vec{l}\cdot 
\vec{\sigma}_1+\vec{q}_2\cdot \vec{\sigma}_1)+q_2^2
\vec{l}\cdot \vec{\sigma}_2\big]+\vec{q}_2\, \big[i \,\vec{l}\cdot 
\vec{\sigma}_1 \vec{l}\cdot \vec{q}_2\times 
\vec{\sigma}_2+\vec{l}\cdot \vec{q}_2 (\vec{l}\cdot 
\vec{\sigma}_1-\vec{l}\cdot \vec{\sigma}_2)+l^2 
(\vec{l}\cdot \vec{\sigma}_1+\vec{q}_2\cdot \vec{\sigma}_2)\big]\nn
&-&i \, \vec{l}\times\vec{q}_2\,
\big[-\vec{l}\cdot \vec{\sigma}_2\, (\vec{l}\cdot 
\vec{\sigma}_1+\vec{q}_2\cdot \vec{\sigma}_1)+(l^2+\vec{l}\cdot \vec{q}_2) 
(\vec{\sigma}_1\cdot \vec{\sigma}_2-1)\big] 
+i \, \vec{l}\times\vec{\sigma}_2\,
\big[l^2 \,\vec{q}_2\cdot \vec{\sigma}_1-\vec{l}\cdot 
\vec{q}_2 \vec{l}\cdot \vec{\sigma}_1\big] \nn
&+&i \, \vec{q}_2
\times\vec{\sigma}_2\,
\big[l^2 \,\vec{q}_2\cdot \vec{\sigma}_1-\vec{l}\cdot 
\vec{q}_2 \vec{l}\cdot \vec{\sigma}_1\big]-\vec{\sigma}_1\,
(l^2+\vec{l}\cdot \vec{q}_2)^2
+\vec{\sigma}_2\,\big[(\vec{l}\cdot \vec{q}_2)^2-l^2 \,q_2^2\big] 
\bigg)~,\\
\vec{\cal A}_{{\rm box} \, (6)}^a&=&
\frac{g_A^5 }{64 F_\pi^4 \omega_l^2 \omega_{\vec{l}+\vec{q}_2}^4}
\big(2 i\, [\tau_1\times\tau_2]^a-3
  \tau_1^a-2 
\tau_2^a\big) \bigg(\vec{l} \,\big[-i\,\vec{l}\cdot \vec{\sigma}_1 
\vec{l}\cdot \vec{q}_2\times \vec{\sigma}_2+ \vec{l}\cdot \vec{q}_2 
(\vec{q}_2\cdot \vec{\sigma}_1-\vec{q}_2\cdot \vec{\sigma}_2)\nn
&+&l^2 \,\vec{q}_2\cdot \vec{\sigma}_1+q_2^2 \,
\vec{l}\cdot \vec{\sigma}_2\big]- \vec{q}_2 \,\big[i \,\vec{l}\cdot 
\vec{\sigma}_1 \vec{l}\cdot \vec{q}_2\times 
\vec{\sigma}_2+\vec{l}\cdot \vec{q}_2 (\vec{l}\cdot 
\vec{\sigma}_1+\vec{l}\cdot \vec{\sigma}_2)+l^2
(\vec{l}\cdot \vec{\sigma}_1-\vec{q}_2\cdot 
\vec{\sigma}_2)\big]\nn
&-&i\,\vec{l}
\times\vec{q}_2\,\big[-\vec{l}\cdot \vec{\sigma}_2 (\vec{l}\cdot 
\vec{\sigma}_1+\vec{q}_2\cdot \vec{\sigma}_1)+(l^2+\vec{l}\cdot \vec{q}_2) 
(\vec{\sigma}_1\cdot \vec{\sigma}_2-1)\big] -i\,
\vec{l}\times\vec{\sigma}_2\,\big[\vec{l}\cdot \vec{q}_2 \vec{l}\cdot 
\vec{\sigma}_1-l^2 \,\vec{q}_2\cdot \vec{\sigma}_1\big]\nn 
&-&i\,\vec{q}_2\times\vec{\sigma}_2\,\big[\vec{l}\cdot \vec{q}_2 \vec{l}\cdot 
\vec{\sigma}_1-l^2 \vec{q}_2\cdot \vec{\sigma}_1\big] 
- \vec{\sigma}_1\,(l^2+\vec{l}\cdot \vec{q}_2
\,)^2 +  \vec{\sigma}_2\,\big[(\vec{l}\cdot 
\vec{q}_2)^2-l^2 \,q_2^2\,\big] 
\bigg)~.
\eeqa
Here, $F_\pi$ denotes the pion decay constant while $\vec \sigma_i$
($\fet \tau_i$) are spin (isospin) Pauli matrices of the nucleon $i$. 
To further simplify these expressions, we write 
the scalar products $l^2 \equiv | \vec l\,|^2$ and $\vec{l}\cdot\vec{q}_2$ as 
linear combinations of the corresponding pion energies via
\beqa
l^2&=&\omega_l^2 - M_\pi^2~, \nn
\vec{l}\cdot\vec{q}_2&=&\frac{1}{2}\big(\omega_{\vec{l}+\vec{q}_2}^2-\omega_l^2-q_2^2\big)~.
\eeqa
After these simplifications we still have to evaluate tensor
integrals up to rank three. To this aim, we carry out the standard Passarino-Veltman
reduction in $d$ dimensions: 
\beqa
&&\mu^{3-d}\int \frac{d^d l}{(2\pi)^d}\frac{l_{i_1} l_{i_2}
  l_{i_3}}{\omega_l^{n_1}\omega_{\vec{l}+\vec{q}_2}^{n_2}} \,=\,
\frac{ (q_2)_{i_3} \delta_{i_1,i_2}+(q_2)_{i_2} 
\delta_{i_1,i_3}+(q_2)_{i_1} \delta_{i_2,i_3}}{8 (d-1) 
q_2^4}\bigg(q_2^4 \left(4 
M_\pi^2+q_2^2\right) s(n_1,n_2)\nn
&-&q_2^2 \big[\left(q_2^2-4 M_\pi^2\right) 
s(n_1-2,n_2)+\left(4 M_\pi^2+3 q_2^2\right) 
s(n_1,n_2-2)+s(n_1-4,n_2)+2 s(n_1-2,n_2-2)\nn
&-& 3 s(n_1,n_2-4)
\big]+s(n_1-6,n_2)-3 s(n_1-4,n_2-2)+3 s(n_1-2,n_2-4)-s(n_1,n_2-6)
\bigg)\nn
&+&\frac{(q_2)_{i_1} (q_2)_{i_2} 
(q_2)_{i_3} }{8 (d-1) q_2^6}\bigg(-q_2^4 s(n_1,n_2) 
\left((d+2) q_2^2+12 M_\pi^2\right)-3 d 
q_2^2 s(n_1-4,n_2)\nn
&+&6 d q_2^2
s(n_1-2,n_2-2)
-3 d q_2^4 s(n_1-2,n_2)-3 d 
q_2^2 s(n_1,n_2-4)+3 d q_2^4 s(n_1,n_2-2)\nn
&-&(d+2) s(n_1-6,n_2)+3 (d+2) s(n_1-4,n_2-2)-3 
d s(n_1-2,n_2-4)+d s(n_1,n_2-6)\nn
&-&12 M_\pi^2 q_2^2 
s(n_1-2,n_2)+12 M_\pi^2 q_2^2 s(n_1,n_2-2)+6 
q_2^2 s(n_1-4,n_2)+6 q_2^4 
s(n_1-2,n_2)\nn
&-&6 q_2^2 s(n_1,n_2-4)+6 
q_2^4 s(n_1,n_2-2)-6 s(n_1-2,n_2-4)+2 
s(n_1,n_2-6)\bigg)~,\\
&&\mu^{3-d}\int \frac{d^d l}{(2\pi)^d}\frac{l_{i_1}
  l_{i_2}}{\omega_l^{n_1}\omega_{\vec{l}+\vec{q}_2}^{n_2}}\,=\,
\frac{(q_2)_{i_1} (q_2)_{i_2} }{4 (d-1) q_2^4}
\bigg(q_2^2 
s(n_1,n_2) \left(d q_2^2+4 M_\pi^2\right)+d \big[2 
q_2^2 (s(n_1-2,n_2)\nn
&-&s(n_1,n_2-2))+s(n_1-4,n_2)-2 
s(n_1-2,n_2-2)+s(n_1,n_2-4)\big]-4 q_2^2 s(n_1-2,n_2)
\bigg)\nn
&-&\frac{\delta_{i_1,i_2} }{4 (d-1) 
q_2^2}\bigg(q_2^2 \big[\left(4 M_\pi^2+ q_2^2
\right) s(n_1,n_2)-2 (s(n_1-2,n_2)+s(n_1,n_2-2))
\big]+s(n_1-4,n_2)\nn
&-&2 s(n_1-2,n_2-2)+s(n_1,n_2-4)\bigg)~,\\
&&\mu^{3-d}\int \frac{d^d
  l}{(2\pi)^d}\frac{l_{i_1}}{\omega_l^{n_1}\omega_{\vec{l}+\vec{q}_2}^{n_2}}
\,=\,
-\frac{(q_2)_{i_1}}{2 q_2^2}(q_2^2 
s(n_1,n_2)+s(n_1-2,n_2)-s(n_1,n_2-2))~,
\eeqa
with the scalar integrals defined by
\beqa
s(n_1,n_2)&=&\mu^{3-d}\int \frac{d^d
  l}{(2\pi)^d}\frac{1}{\omega_l^{n_1}\omega_{\vec{l}+\vec{q}_2}^{n_2}}~.
\eeqa
where $\mu$ denotes the scale of dimensional regularization. 
The scalar integrals can be further simplified. First, if both of the indices
$n_1$ and $n_2$ are negative or zero, the function $s(n_1,n_2)$
vanishes in dimensional regularization:
\beqa
s(n_1,n_2)&=&0\quad{\rm for}\quad n_1\leq 0 \;\; \&\;\;  n_2\leq 0~.
\eeqa
Further, the scalar integrals are symmetric
\beqa
s(n_1,n_2)&=&s(n_2,n_1)~,
\eeqa
as follows trivially from the substitution
$\vec{l}\to-\vec{l}-\vec{q}_2$. If one of the indices $n_1$ or $n_2$
equals  zero, the
tensor integrals can be simplified to
\beqa
&&\int \frac{d^d l}{(2\pi)^d}\frac{l_{i_1} l_{i_2}
  l_{i_3}}{\omega_l^{n_1}}\,=\, \int \frac{d^d l}{(2\pi)^d}\frac{l_{i_1} }{\omega_l^{n_1}} \,=\,0~,\nn
&&\mu^{3-d}\int \frac{d^d l}{(2\pi)^d}\frac{l_{i_1}
  l_{i_2}}{\omega_l^{n_1}}\,=\,
\frac{\delta_{i_1,i_2}}{d}\big(s(n_1-2,0)- M_\pi^2 s(n_1,0)\big)~.\label{tadpole_reduction}
\eeqa
For negative indices, in particular $-5<n_2<0$, we can use Eq.~(\ref{tadpole_reduction}) to
reduce the indices to zero. The scalar integrals of this kind 
needed for this calculation are given by
\beqa
s(4,-4)&=&
\frac{q_2^2 }{d}\left(s(4,0) \left(d\, q_2^2-4 M_\pi^2\right)+2 
(d+2) s(2,0)\right)~,\\
s(4,-2)&=&q_2^2 s(4,0)+s(2,0)~.
\eeqa
To further reduce the scalar integrals we use the partial integration
technique
\beqa
\int \frac{d^d l}{(2\pi)^d}\frac{\partial}{\partial
  \vec{l}}\cdot\vec{X}\frac{1}{\omega_l^{n_1}
\omega_{\vec{l}+\vec{q}_2}^{n_2}}&=&0~,\label{partial_int}
\eeqa
where $\vec{X}$ represents $\vec{l}$ or
$\vec{q}_2$. Eq.~(\ref{partial_int}) leads to the following reduction formula for
the scalar integrals:
\beqa
s(n_1,n_2)&=&
\frac{1}{(n_2-2) q_2^2 \left(4 M_\pi^2+q_2^2\right)}\bigg(s(n_1,n_2-2)
\left(q_2^2 (-2 d+2 n_1+n_2-2)+2 M_\pi^2 (n_1-n_2+2)
\right)\label{sn1n2:reduction}\\
&+&(n_2-2) \left(2 M_\pi^2+q_2^2\right) s(n_1-2,n_2)-2 M_\pi^2 n_1 
s(n_1+2,n_2-4)\bigg)~, \nn
s(n_1,0)&=&-\frac{(d-n_1+2) s(n_1-2,0)}{M_\pi^2 (n_1-2)}~. \label{sn1:reduction}
\eeqa
Eqs.~(\ref{sn1n2:reduction}) and (\ref{sn1:reduction}) are used to
express all scalar integrals in $3$ dimensions in terms of 
\beqa
s(2,0)&=&-\frac{M_\pi}{4\pi}~,\quad
s(2,2)\,=\,\frac{A(q_2)}{2\pi}~,
\eeqa
where 
\beqa
A(q_2)&=&\frac{1}{2 q_2}\arctan\bigg(\frac{q_2}{2 M_\pi}\bigg)~.
\eeqa

Having performed the Passarino-Veltman reduction as described above,
the expressions for box-diagrams of Fig.~\ref{fig:boxdiagr} in $d=3$ dimensions simplify to
\beqa
\vec{\cal A}_{{\rm box} \, (1)}^a&=&
\frac{g_A^5 }{512 \pi  
F_\pi^4}\left(2 i
    [\tau_1\times\tau_2]^a-3 \tau_1^a+2 \tau_2^a\right)
 \bigg\{-A(q_2) \left(q_2^2 
\left(\vec{\sigma}_2+i \vec{\sigma}_1\times\vec{\sigma}_2\right)+i 
\vec{q}_2\cdot \vec{\sigma}_2 \vec{q}_2\times\vec{\sigma}_1
\right)\nn
&-&\left(3 M_\pi +2 A(q_2) \left(2 M_\pi^2+q_2^2\right)-\frac{M_\pi^3}{4 
M_\pi^2+q_2^2}\right) \vec{\sigma}_1+A(q_2) \vec{q}_2\cdot \vec{\sigma}_2
\vec{q}_2\bigg\}\;,\\
\vec{\cal A}_{{\rm box} \, (2)}^a&=&
\frac{g_A^5}{512 \pi  F_\pi^4} \left(2 \tau_2^a-\tau_1^a\right) \bigg\{\left(3 M_\pi +q_2^2 A(q_2)-\frac{2 M_\pi^3}{4 M_\pi^2+q_2^2}\right) 
\vec{\sigma}_1+2 q_2^2 A(q_2) \vec{\sigma}_2\nn
&+&\vec{q}_2 
\bigg(\bigg[\frac{A(q_2) \left(8 M_\pi^2+q_2^2\right)-2 M_\pi 
}{q_2^2}-\frac{M_\pi}{4 M_\pi^2+q_2^2}\bigg]\vec{q}_2\cdot \vec{\sigma}_1-2 A(q_2) \vec{q}_2\cdot 
\vec{\sigma}_2
\bigg)\bigg\}\;,\\
\vec{\cal A}_{{\rm box} \, (3)}^a&=&
\frac{g_A^5 }{512 \pi  F_\pi^4}\left(2 i
  [\tau_1\times\tau_2]^a+3 \tau_1^a-2 \tau_2^a
\right) \bigg\{A(q_2) \left(q_2^2 \left(\vec{\sigma}_2-i 
\vec{\sigma}_1\times\vec{\sigma}_2\right)-i\vec{q}_2\cdot \vec{
\sigma}_2 \vec{q}_2\times\vec{\sigma}_1\right)\nn
&+&\left(3 M_\pi+2 A(q_2) \left(2 M_\pi^2+q_2^2
\right)-\frac{M_\pi^3}{4 M_\pi^2+q_2^2}\right) 
\vec{\sigma}_1 -A(q_2) 
\vec{q}_2\cdot \vec{\sigma}_2 \vec{q}_2\bigg\}\;,\\
\vec{\cal A}_{{\rm box} \, (4)}^a&=&
\frac{g_A^5 }{512 \pi  F_\pi^4}\left(2 i [\tau_1\times\tau_2]^a +3 \tau_1^a+2 \tau_2^a
\right) \bigg\{A(q_2) \left(q_2^2 \left(\vec{\sigma}_2+i 
\vec{\sigma}_1\times\vec{\sigma}_2\right)+i \vec{q}_2\cdot 
\vec{\sigma}_2 \vec{q}_2\times\vec{\sigma}_1\right)\nn
&-&\left(3 M_\pi +2 A(q_2) \left(2 M_\pi^2+q_2^2
\right)-\frac{M_\pi^3}{4 M_\pi^2+q_2^2}\right) 
\vec{\sigma}_1-A(q_2) 
\vec{q}_2\cdot \vec{\sigma}_2 \vec{q}_2\bigg\}\;,\\
\vec{\cal A}_{{\rm box} \, (5)}^a&=&
\frac{g_A^5 }{512 \pi  F_\pi^4} \left(2 \tau_2^a+\tau_1^a\right) \bigg\{\left(3 M_\pi+q_2^2 A(q_2)-\frac{2 M_\pi^3}{4 
M_\pi^2+q_2^2}\right) \vec{\sigma}_1 -2 q_2^2 A(q_2) 
\vec{\sigma}_2\nn
&+&\vec{q}_2 \bigg(
\bigg[\frac{A(q_2) \left(8 M_\pi^2+q_2^2\right)-2 M_\pi 
}{q_2^2}-\frac{M_\pi}{4 
M_\pi^2+q_2^2}\bigg]\vec{q}_2\cdot \vec{\sigma}_1+2 A(q_2) \vec{q}_2\cdot 
\vec{\sigma}_2\bigg)\bigg\}\;,\\
\vec{\cal A}_{{\rm box} \, (6)}^a&=&
\frac{g_A^5 }{512 \pi  F_\pi^4} \left(2 i [\tau_1\times\tau_2]^a-3 \tau_1^a-2 \tau_2^a
\right) \bigg\{-A(q_2) \left(q_2^2 
\left(\vec{\sigma}_2-i \vec{\sigma}_1\times\vec{\sigma}_2\right)-i 
\vec{q}_2\cdot \vec{\sigma}_2 \vec{q}_2\times\vec{\sigma}_1
\right)\nn
&+&\left(3 M_\pi +2 A(q_2) \left(2 M_\pi^2+q_2^2
\right)-\frac{M_\pi^3}{4 M_\pi^2+q_2^2}\right) 
\vec{\sigma}_1+A(q_2) \vec{q}_2\cdot \vec{\sigma}_2 \vec{q}_2
\bigg\}\;.
\eeqa
Our final result for the sum of all six box diagrams takes the form 
\beqa
\label{final}
\vec{\cal A}_{{\rm box}}^a&=&\frac{g_A^5}{128 \pi F_\pi^4} \bigg\{
\tau_2^a \bigg[ \vec{q}_2\cdot \vec{\sigma}_1 \vec{q}_2 \bigg(
\frac{(8 M_\pi^2+q_2^2) A(q_2) - 2 M_\pi}{q_2^2} - \frac{M_\pi}{4
  M_\pi^2 + q_2^2} \bigg) - \vec \sigma_1 \Big((8 M_\pi^2+3 q_2^2)
A(q_2) + 3 M_\pi \Big) \bigg] \nn
&+& 2 \tau_1^a \Big( \vec \sigma_2 q_2^2 - \vec{q}_2\cdot
\vec{\sigma}_2 \vec{q}_2 \Big) A(q_2) \bigg\}\; +\;  (1\,\leftrightarrow\,2)~,
\eeqa
with $q_2 \equiv |\vec q_2 |$.
This expression coincides with the $g_A^5$-terms in Eqs.~(5.29) and
(5.31) of our paper \cite{Krebs:2016rqz}
when we set $\vec{k}=0$ to switch off all pion-pole contributions, but
differs from the corresponding term in Eq.~(7.5) of Ref.~\cite{Baroni:2015uza}.

\section{Summary and conclusions}
\label{sec4}
\label{sec:summary}
In this paper we have re-derived the non-pion-pole box-diagram
contribution to the two-pion-exchange axial-vector current operator
proportional to $g_A^5$ using the approach of the JLab-Pisa group 
\cite{Baroni:2015uza,Baroni:2016xll,Pastore:2011ip}.
We have shown at the Fock-space level that 
the axial-vector currents, constructed using the method of unitary
transformation \cite{Krebs:2016rqz} and the time-ordered perturbation theory
approach of Ref.~\cite{Baroni:2015uza}, are unitary
equivalent. The off-shell conditions employed by the JLab-Pisa group
are found to correspond to the following choice of the unitary
phases 
\beqa
\alpha_i^{ax}&=&0, \quad \alpha_1\,=\,-\frac{1}{2}, \quad
\alpha_2\,=\,\frac{1}{4}\,,
\eeqa
in the notation of Ref.~\cite{Krebs:2016rqz}. 
Although the phases $\alpha_i^{ax}$ are chosen differently by our
group in order to enforce renormalizability of the current operators, the
two-pion-exchange contributions proportional to $g_A^5$ are
unaffected by this choice\footnote{The two-pion exchange contributions
to the axial current $\propto g_A^5$ only depend on the phase $\alpha_1^{ax}$, which
has been set to zero in Ref.~\cite{Krebs:2016rqz}.} and should
therefore coincide in both approaches.
We have presented a detailed calculation of the box
diagrams starting from the corresponding Fock-space operators and including
intermediate steps, thereby reproducing our original result from
Ref.~\cite{Krebs:2016rqz}. While we have not been able to identify the
origin of the disagreement between the two results, see
Refs.~\cite{Krebs:2016rqz,Baroni:2018fdn}, we hope that our
work will help to resolve this issue in the future. 
\section*{Acknowledgments}

This work is supported in part by  the DFG (Grant No. TRR110)
and the NSFC (Grant No. 11621131001) through the funds provided
to the Sino-German CRC 110 ``Symmetries and the Emergence of
Structure in QCD", by the BMBF (Grant No. 05P18PCFP1), 
by the Chinese  Academy of Sciences (CAS) President's International Fellowship Initiative (PIFI)
(grant no. 2018DM0034)  and by VolkswagenStiftung (grant no. 93562).


\appendix
\numberwithin{equation}{section}
\section{Transfer matrix with time-dependent interaction}
\label{app:transfer_matrix}
In this appendix we consider the nuclear transfer matrix 
in the presence of explicitly time-dependent
interactions involving external classical sources. 
We start with the Schr\"odinger equation
\beqa
\bigg(i\,\frac{\partial}{\partial t}-H_0\bigg)|\Psi(t)\rangle&=& V(t) |\Psi(t)\rangle.
\eeqa
As usual, we introduce a free retarded Green's function which satisfies
\beqa
\bigg(i\frac{\partial}{\partial t}-H_0\bigg)
G_+(t-t^\prime)&=&\delta (t-t^\prime)\quad{\rm and}\quad
G_+(t-t^\prime)\,=\,0\quad {\rm for}\quad t<t^\prime\;,
\eeqa
and is given by
\beqa
G_+(t-t^\prime)&=&-i\,\theta(t-t^\prime)\exp\big[-i\,\big(H_0 - i\,\epsilon\big)\,(t-t^\prime)\big].
\eeqa
The solution of the Schr\"odinger equation can be written as
\beqa
|\Psi^+(t)\rangle&=&|\phi(t)\rangle + \int_{-\infty}^{\infty} d
t^\prime \, G_+(t-t^\prime) V(t^\prime)|\Psi^+(t^\prime)\rangle\;,\label{solution_schroedinger}
\eeqa
with the state $|\phi(t)\rangle$ satisfying the free Schr\"odinger
equation
\beqa
\bigg(i\,\frac{\partial}{\partial t} - H_0\bigg)|\phi(t)\rangle &=&0.
\eeqa
We now take the Fourier transform of Eq.~(\ref{solution_schroedinger})
by multiplying both sides with $e^{i E t}$ and integrating over time:
\beqa
|\tilde\Psi^+(E)\rangle&=&|\tilde\phi(E)\rangle + \int_{-\infty}^{\infty} d t\, d
t^\prime \,e^{i\, E t}G_+(t - t^\prime) V(t^\prime) |\Psi^+(t^\prime)\rangle,\label{fourier_tr:schroedinger_eq}
\eeqa
where we have defined 
\beqa
|\tilde\Psi^+(E)\rangle&=&\int_{-\infty}^{\infty} d t \, e^{i E t}
|\Psi^+(t)\rangle, \quad \quad |\tilde\phi(E)\rangle\,=\, \int_{-\infty}^{\infty} d t \, e^{i E t} |\phi(t)\rangle.
\eeqa
To simplify Eq.~(\ref{fourier_tr:schroedinger_eq}), we Fourier-transform the
free Green's function $G_+$: 
\beqa
\tilde G_+(E)&=&\int_{-\infty}^\infty d t\, e^{i E t} G_+(t)\,=\,
-i\,\int_{0}^\infty d t \,e^{i (E - H_0+ i\,\epsilon)
  t}\,=\,\frac{1}{E - H_0 + i\,\epsilon}.
\eeqa
The backwards transformations are given by
\beqa
G_+(t)&=&\int \frac{d E}{2\pi} e^{-i\, E t}\tilde G_+(E), \quad \quad
|\Psi^+(t)\rangle\,=\,\int \frac{d E}{2\pi} e^{-i\, E t}|\tilde \Psi^+(E)\rangle\;.\label{fourier_back}
\eeqa
Using Eq.~(\ref{fourier_back}), we can rewrite
Eq.~(\ref{fourier_tr:schroedinger_eq}) into
\beqa
|\tilde\Psi^+(E)\rangle&=&|\tilde\phi(E)\rangle + \tilde G_+(E)\int
\frac{d E^\prime}{2\pi}\tilde V(E -
E^\prime)|\tilde\Psi^+(E^\prime)\rangle\label{tmatrix:prepare}\\
&=&|\tilde\phi(E)\rangle + \tilde G_+(E) \int \frac{d E^\prime}{2\pi} T(E,E^\prime)|\phi(E^\prime)\rangle.\label{tmatrix:definition}
\eeqa
Equation~({\ref{tmatrix:definition}}) defines the transfer matrix in the
presence of an external source. It satisfies the integral equation
\beqa
T(E,E^\prime)&=&\tilde V(E-E^\prime) + \int \frac{d E^{\prime\prime}}{2\pi} \tilde
V(E-E^{\prime\prime})\tilde G_+(E^{\prime\prime}) T(E^{\prime\prime}, E^\prime),\label{transition_int_eq}
\eeqa
which can also be written in the equivalent form 
\beqa
T(E,E^\prime)&=&\tilde V(E-E^\prime) + \int \frac{d E^{\prime\prime}}{2\pi} T(E, E^{\prime\prime}) \tilde G_+(E^{\prime\prime})\tilde
V(E^{\prime\prime}-E^\prime).\label{transition_int_eq:transpose}
\eeqa
Rewriting Eq.~(\ref{transition_int_eq:transpose}) into 
\beqa
\int \frac{d E^{\prime\prime}}{2\pi} T(E, E^{\prime\prime})\Big(2\pi\delta
(E^{\prime\prime}-E^\prime) -
\tilde G_+(E^{\prime\prime})\tilde
V(E^{\prime\prime}-E^\prime)\Big)&=&\tilde V(E-E^\prime), 
\label{inverse_tmatrix}
\eeqa
and replacing $\tilde V(E-E^\prime )$ in Eq.~(\ref{tmatrix:prepare})
by left hand side of Eq.~(\ref{inverse_tmatrix}) we obtain
\beqa
|\tilde\Psi^+(E)\rangle&=&|\tilde \phi(E)\rangle  + \tilde G_+(E)\int
\frac{d E^\prime}{2\pi} \int \frac{d E^{\prime\prime}}{2\pi}
T(E,E^{\prime\prime})\Big(2\pi \delta(E^{\prime\prime}-E^\prime) - \tilde G_+(E^{\prime\prime})\tilde
V(E^{\prime\prime}-E^\prime)\Big)|\tilde \Psi^+(E^\prime)\rangle.\quad
\eeqa
Using 
\beqa
\int \frac{d E^{\prime}}{2\pi}\Big(2\pi \delta
  (E^{\prime\prime}-E^\prime) - \tilde G_+(E^{\prime\prime})\tilde
  V(E^{\prime\prime}-E^\prime)\Big)|\tilde\Psi^+(E^\prime)\rangle&=&|\tilde \phi(E^{\prime\prime})\rangle
\eeqa
we indeed obtain Eq.~(\ref{tmatrix:definition}). 

As a next step, we decompose the interaction into the time-dependent and
time-independent parts
\beqa
V(E-E^\prime)&=&2\pi \delta(E-E^\prime) v + v_5(E-E^\prime),
\eeqa
where $v$ denotes the time-independent nuclear potential while $v_5$ is 
the part of the interaction that depends on the 
external axial-vector source. In a similar way, the
transfer matrix can be decomposed as
\beqa
T(E,E^\prime)&=&2\pi \delta(E-E^\prime) t(E) + t_5(E,E^\prime)\;.
\eeqa
The off-shell transfer matrix $t(E)$ satisfies the usual Lippmann-Schwinger equation
\beqa
t(E)&=&v + v\, \tilde G_+(E) \,t(E),
\eeqa 
while the transfer matrix $t_5$ fulfills the integral equation 
\beqa
t_5(E,E^\prime)&=&
v_5(E-E^\prime) + v\tilde \, G_+(E) \, t_5(E, E^\prime) + v_5
\, (E-E^\prime)\, \tilde G_+(E^\prime) \, t(E^\prime) \nn
&+& \int \frac{d
  E^{\prime\prime}}{2\pi} \, v_5(E-E^{\prime\prime}) \, \tilde
G_+(E^{\prime\prime}) \, t_5(E^{\prime\prime},E^\prime).
\eeqa
The last term contributes only to processes with at least two external
sources and is, therefore, not considered in our calculation. Taking
into account only processes with at most a single insertion of the
external sources, we obtain for the transfer matrix
\beqa
\big(1-v \tilde G_+(E)\big) t_5(E, E^\prime) &=& v_5(E-E^\prime)\big(1
+ \tilde G_+(E^\prime) t(E^\prime)\big)\nn
&=&v_5(E-E^\prime)\big(1+\tilde G_+(E^\prime) v\big(1-\tilde
G_+(E^\prime) v\big)^{-1}\big)\nn
&=&v_5(E-E^\prime) \Big(1-\tilde G_+(E^\prime) v+\tilde
G_+(E^\prime) v\Big)\big(1-\tilde G_+(E^\prime) v\big)^{-1}\nn
&=&v_5(E-E^\prime) \big(1-\tilde G_+(E^\prime) v\big)^{-1}
\eeqa
which leads to the final expression
\beqa
t_5(E, E^\prime)&=& \big(1-v \tilde
G_+(E)\big)^{-1}v_5(E-E^\prime) 
\big(1-\tilde G_+(E^\prime) v\big)^{-1}.
\eeqa
One observes, in particular, that all energies in the operator on the left-hand (right-hand)
side of $v_5$ correspond to the 
final (initial) state energies. Last but not least, we emphasize that
the derived integral equations also follow from the well-known
two-potential formalism in scattering theory.  
\section{Fock-space expressions for the axial-vector operator}
\label{app:axial_vector_fock_space}
In this appendix we give the expressions for the axial-vector current operator
in Fock space. The corresponding Mathematica notebook is available
upon request from the authors. As explained in section
\ref{sec2}, the axial-vector current $v_5^{(n)}$ at the chiral order $Q^n$
is obtained following the approach or Refs.~\cite{Baroni:2015uza,Baroni:2016xll} by inverting
Eq.~(\ref{v5_FD_inversion}) for the transfer matrix. We find the following results at the
first four orders:  
\beqa
v_5^{(-3)}&=&\frac{\eta A_N \eta}{2}-\eta A_\pi 
\frac{\lambda^1}{\omega} V \eta+ \;{\rm h.c.}\;,\\
v_5^{(-2)}&=&\eta A_\pi {\cal E} \frac{\lambda^1}{\omega^2} V 
\eta-\eta H_0 \eta V \frac{\lambda^1}{\omega^2} 
A_\pi \eta + \;{\rm h.c.}\;,\\
v_5^{(-1)}&=&2\eta H_0 \eta V {\cal E} \frac{\lambda^1}{\omega^3} 
A_\pi \eta
-\eta H_0^2 \eta V \frac{\lambda^1}{\omega^3}
A_\pi \eta
-\frac{1}{2}\eta A_N \eta V \frac{\lambda^1}{\omega^2} 
V \eta
+\frac{1}{2}\eta V \frac{\lambda^1}{\omega} A_N \frac{\lambda^1}{\omega} 
V \eta
-\eta A_\pi {\cal E}^2 
\frac{\lambda^1}{\omega^3} V \eta\nn
&+&\eta A_\pi \frac{\lambda^1}{\omega^2} V \eta V 
\frac{\lambda^1}{\omega} V \eta
+\frac{1}{2}\eta A_\pi \frac{\lambda^1}{\omega} V \eta 
V \frac{\lambda^1}{\omega^2} V \eta
-\eta A_\pi \frac{\lambda^1}{\omega} V 
\frac{\lambda^2}{\omega} V 
\frac{\lambda^1}{\omega} V \eta
+\frac{1}{2}\eta V \frac{\lambda^1}{\omega^2} V \eta 
A_\pi \frac{\lambda^1}{\omega} V \eta\nn
&-&\eta 
V \frac{\lambda^1}{\omega} A_\pi 
\frac{\lambda^2}{\omega} V 
\frac{\lambda^1}{\omega} V \eta+ \;{\rm h.c.}\;,\\
v_5^{(0)}&=&-\frac{1}{2}\eta H_0 \eta V \frac{\lambda^1}{\omega^3} 
V \eta A_N \eta+\eta H_0 \eta V 
\frac{\lambda^1}{\omega^2} A_N \frac{\lambda^1}{\omega} 
V \eta
-\frac{1}{2}\eta A_N \eta H_0 \eta 
V \frac{\lambda^1}{\omega^3} V 
\eta
+\frac{1}{2}\eta H_0 \eta V 
\frac{\lambda^1}{\omega^3} V \eta A_\pi 
\frac{\lambda^1}{\omega} V \eta\nn
&+&\frac{1}{2}\eta H_0 \eta 
V \frac{\lambda^1}{\omega^3} V \eta 
V \frac{\lambda^1}{\omega} A_\pi 
\eta+\frac{1}{2}\eta H_0 \eta V 
\frac{\lambda^1}{\omega^2} A_\pi \eta V 
\frac{\lambda^1}{\omega^2} V \eta
-\eta H_0 \eta V \frac{\lambda^1}{\omega^2} A_\pi 
\frac{\lambda^2}{\omega} V 
\frac{\lambda^1}{\omega} V \eta
+\eta H_0 \eta 
V \frac{\lambda^1}{\omega^2} V \eta 
V \frac{\lambda^1}{\omega^2} A_\pi \eta\nn
&-&\eta H_0 \eta V \frac{\lambda^1}{\omega^2} V 
\frac{\lambda^2}{\omega} A_\pi 
\frac{\lambda^1}{\omega} V \eta-\eta H_0 \eta 
V \frac{\lambda^1}{\omega^2} V 
\frac{\lambda^2}{\omega} V 
\frac{\lambda^1}{\omega} A_\pi \eta
+\eta H_0 \eta V \frac{\lambda^1}{\omega} V \eta 
V \frac{\lambda^1}{\omega^3} A_\pi \eta
-\eta H_0 \eta V \frac{\lambda^1}{\omega} V 
\frac{\lambda^2}{\omega^2} A_\pi 
\frac{\lambda^1}{\omega} V \eta\nn
&-&\eta H_0 \eta V \frac{\lambda^1}{\omega} V 
\frac{\lambda^2}{\omega^2} V 
\frac{\lambda^1}{\omega} A_\pi \eta
-\eta H_0 \eta 
V \frac{\lambda^1}{\omega} V 
\frac{\lambda^2}{\omega} V 
\frac{\lambda^1}{\omega^2} A_\pi \eta
+\eta A_\pi \frac{\lambda^1}{\omega^3} V \eta H_0 \eta V 
\frac{\lambda^1}{\omega} V \eta+\frac{1}{2}\eta 
A_\pi \frac{\lambda^1}{\omega^2} V \eta H_0 \eta 
V \frac{\lambda^1}{\omega^2} V 
\eta\nn
&+&\frac{1}{2}\eta A_\pi \frac{\lambda^1}{\omega} 
V \eta H_0 \eta V \frac{\lambda^1}{\omega^3} 
V \eta+\frac{1}{2}\eta V 
\frac{\lambda^1}{\omega^3} V \eta H_0 \eta A_\pi 
\frac{\lambda^1}{\omega} V \eta+\eta A_N \eta 
V {\cal E} \frac{\lambda^1}{\omega^3} V \eta
-\eta V {\cal E} \frac{\lambda^1}{\omega^2} A_N 
\frac{\lambda^1}{\omega} V \eta\nn
&-&2\eta A_\pi {\cal E} 
\frac{\lambda^1}{\omega^3} V \eta V 
\frac{\lambda^1}{\omega} V \eta-\frac{1}{2}\eta 
A_\pi {\cal E} \frac{\lambda^1}{\omega^2} V \eta 
V \frac{\lambda^1}{\omega^2} V \eta+\eta 
A_\pi {\cal E} \frac{\lambda^1}{\omega^2} V 
\frac{\lambda^2}{\omega} V 
\frac{\lambda^1}{\omega} V \eta-\eta A_\pi 
\frac{\lambda^1}{\omega^2} V \eta V {\cal E} 
\frac{\lambda^1}{\omega^2} V \eta\nn
&-&\eta A_\pi 
\frac{\lambda^1}{\omega} V \eta V {\cal E} 
\frac{\lambda^1}{\omega^3} V \eta+\eta A_\pi 
\frac{\lambda^1}{\omega} V {\cal E} 
\frac{\lambda^2}{\omega^2} V 
\frac{\lambda^1}{\omega} V \eta+\eta A_\pi 
\frac{\lambda^1}{\omega} V 
\frac{\lambda^2}{\omega} V {\cal E} 
\frac{\lambda^1}{\omega^2} V \eta
-\eta V {\cal E} 
\frac{\lambda^1}{\omega^3} V \eta A_\pi 
\frac{\lambda^1}{\omega} V \eta\nn
&-&\frac{1}{2}\eta 
V {\cal E} \frac{\lambda^1}{\omega^2} A_\pi \eta 
V \frac{\lambda^1}{\omega^2} V \eta+\eta 
V {\cal E} \frac{\lambda^1}{\omega^2} A_\pi 
\frac{\lambda^2}{\omega} V 
\frac{\lambda^1}{\omega} V \eta+\eta V {\cal E} 
\frac{\lambda^1}{\omega^2} V 
\frac{\lambda^2}{\omega} A_\pi 
\frac{\lambda^1}{\omega} V \eta+\eta V \frac{
\lambda^1}{\omega} A_\pi {\cal E} 
\frac{\lambda^2}{\omega^2} V 
\frac{\lambda^1}{\omega} V \eta\nn
&+& \;{\rm h.c.}\;.
\eeqa
Here, $A_\pi$  ($A_N \propto g_A$) refer to the lowest-order vertices
that describe the coupling of the external axial source to a single pion field (two nucleon
fields). In the notation of Ref.~\cite{Krebs:2016rqz}, $A_N$
and $A_\pi$ correspond to the Fock-space operators $A_{2,0}^{(0)}$ and
$A_{0,1}^{(-1)}$, respectively.    
Since the expressions for $v_5^{(1)}$ are rather lengthy, we retain
below only those contributions which are relevant for the box diagrams,
$v_{5\rm b}^{(1)}$, i.e.~the terms  $\propto g_A^5$.
Further, we distinguish between the pion-pole
contributions  driven by $A_\pi$, $v_{5\rm b, \, \rm p}^{(1)}$, and
the non-pion-pole terms $v_{5\rm b, \, \rm np}^{(1)}$ associated with $A_N$:
\beqa
v_{5\rm b}^{(1)}&=&v_{5\rm b, \, \rm p}^{(1)}+v_{5\rm b, \, \rm np}^{(1)}.
\eeqa
The expressions for the non-pion-pole contributions have the form:
\beqa
v_{5\rm b, \, \rm np}^{(1)}&=& 
\alpha_1\bigg(\eta A_N \eta V 
\frac{\lambda^1}{\omega} V \eta V 
\frac{\lambda^1}{\omega^3} V \eta-\eta A_N \eta 
V \frac{\lambda^1}{\omega^3} V \eta 
V \frac{\lambda^1}{\omega} V 
\eta\bigg)\nn
&+&\alpha_2\bigg(\eta A_N \eta V 
\frac{\lambda^1}{\omega} V 
\frac{\lambda^2}{\omega} V 
\frac{\lambda^1}{\omega^2} V \eta-\eta A_N \eta 
V \frac{\lambda^1}{\omega^2} V 
\frac{\lambda^2}{\omega} V 
\frac{\lambda^1}{\omega} V \eta\bigg)\nn
&+&\frac{1}{2}\eta 
A_N \eta V \frac{\lambda^1}{\omega^3} 
V \eta V \frac{\lambda^1}{\omega} 
V \eta+\frac{3}{8}\eta A_N \eta V 
\frac{\lambda^1}{\omega^2} V \eta V 
\frac{\lambda^1}{\omega^2} V \eta
-\frac{1}{2}\eta 
A_N \eta V \frac{\lambda^1}{\omega^2} 
V \frac{\lambda^2}{\omega} V 
\frac{\lambda^1}{\omega} V \eta+\frac{1}{2}\eta 
A_N \eta V \frac{\lambda^1}{\omega} 
V \eta V \frac{\lambda^1}{\omega^3} 
V \eta\nn
&-&\frac{1}{2}\eta A_N \eta V 
\frac{\lambda^1}{\omega} V 
\frac{\lambda^2}{\omega^2} V 
\frac{\lambda^1}{\omega} V \eta-\frac{1}{2}\eta 
A_N \eta V \frac{\lambda^1}{\omega} 
V \frac{\lambda^2}{\omega} V 
\frac{\lambda^1}{\omega^2} V \eta
+\frac{1}{8}\eta 
V \frac{\lambda^1}{\omega^2} V \eta 
A_N \eta V \frac{\lambda^1}{\omega^2} 
V \eta\nn
&-&\frac{1}{2}\eta V 
\frac{\lambda^1}{\omega^2} V \eta V 
\frac{\lambda^1}{\omega} A_N \frac{\lambda^1}{\omega} 
V \eta-\eta V \frac{\lambda^1}{\omega} 
A_N \frac{\lambda^1}{\omega^2} V \eta 
V \frac{\lambda^1}{\omega} V \eta+\eta 
V \frac{\lambda^1}{\omega} A_N 
\frac{\lambda^1}{\omega} V 
\frac{\lambda^2}{\omega} V 
\frac{\lambda^1}{\omega} V \eta\nn
&+&\frac{1}{2}\eta 
V \frac{\lambda^1}{\omega} V 
\frac{\lambda^2}{\omega} A_N 
\frac{\lambda^2}{\omega} V 
\frac{\lambda^1}{\omega} V \eta+ \;{\rm h.c.}\;.
\label{NP}
\eeqa
These are the operators needed to derive the expression for the axial
vector current in Eq.~(\ref{final}) at the vanishing momentum of the
external source.
For the sake of completeness, we also give  the operators
leading the pion-pole contributions:
\beqa
v_{5\rm b, \, \rm p}^{(1)}&=&
\alpha_1\bigg(\eta A_\pi \frac{\lambda^1}{\omega} 
V \eta V \frac{\lambda^1}{\omega^3} 
V \eta V \frac{\lambda^1}{\omega} 
V \eta-\eta A_\pi \frac{\lambda^1}{\omega} 
V \eta V \frac{\lambda^1}{\omega} 
V \eta V \frac{\lambda^1}{\omega^3} 
V \eta-\eta V \frac{\lambda^1}{\omega^3} 
V \eta V \frac{\lambda^1}{\omega} 
V \eta A_\pi \frac{\lambda^1}{\omega} 
V \eta\nn
&+&\eta V \frac{\lambda^1}{\omega} 
A_\pi \eta V \frac{\lambda^1}{\omega^3} 
V \eta V \frac{\lambda^1}{\omega} 
V \eta\bigg)
+\alpha_2\bigg(\eta 
A_\pi \frac{\lambda^1}{\omega} V \eta 
V \frac{\lambda^1}{\omega^2} V 
\frac{\lambda^2}{\omega} V 
\frac{\lambda^1}{\omega} V \eta-\eta A_\pi 
\frac{\lambda^1}{\omega} V \eta V 
\frac{\lambda^1}{\omega} V 
\frac{\lambda^2}{\omega} V 
\frac{\lambda^1}{\omega^2} V \eta\nn
&-&\eta V 
\frac{\lambda^1}{\omega^2} V 
\frac{\lambda^2}{\omega} V 
\frac{\lambda^1}{\omega} V \eta A_\pi 
\frac{\lambda^1}{\omega} V \eta+\eta V \frac{
\lambda^1}{\omega} A_\pi \eta V 
\frac{\lambda^1}{\omega^2} V 
\frac{\lambda^2}{\omega} V 
\frac{\lambda^1}{\omega} V \eta\bigg)
-\eta A_\pi \frac{\lambda^1}{\omega^3} V \eta 
V \frac{\lambda^1}{\omega} V \eta 
V \frac{\lambda^1}{\omega} V \eta \nn
&-&\eta A_\pi \frac{\lambda^1}{\omega^2} V \eta 
V \frac{\lambda^1}{\omega^2} V \eta 
V \frac{\lambda^1}{\omega} V \eta-\frac{1}{2}
\eta A_\pi \frac{\lambda^1}{\omega^2} V \eta 
V \frac{\lambda^1}{\omega} V \eta 
V \frac{\lambda^1}{\omega^2} V \eta+\eta 
A_\pi \frac{\lambda^1}{\omega^2} V \eta 
V \frac{\lambda^1}{\omega} V 
\frac{\lambda^2}{\omega} V 
\frac{\lambda^1}{\omega} V \eta\nn
&+&\eta A_\pi 
\frac{\lambda^1}{\omega^2} V 
\frac{\lambda^2}{\omega} V 
\frac{\lambda^1}{\omega} V \eta V 
\frac{\lambda^1}{\omega} V \eta-\frac{1}{2}\eta 
A_\pi \frac{\lambda^1}{\omega} V \eta 
V \frac{\lambda^1}{\omega^3} V \eta 
V \frac{\lambda^1}{\omega} V \eta-\frac{3}{8}
\eta A_\pi \frac{\lambda^1}{\omega} V \eta 
V \frac{\lambda^1}{\omega^2} V \eta 
V \frac{\lambda^1}{\omega^2} V 
\eta\nn
&+&\frac{1}{2}\eta A_\pi \frac{\lambda^1}{\omega} 
V \eta V \frac{\lambda^1}{\omega^2} 
V \frac{\lambda^2}{\omega} V 
\frac{\lambda^1}{\omega} V \eta-\frac{1}{2}\eta 
A_\pi \frac{\lambda^1}{\omega} V \eta 
V \frac{\lambda^1}{\omega} V \eta 
V \frac{\lambda^1}{\omega^3} V 
\eta+\frac{1}{2}\eta A_\pi \frac{\lambda^1}{\omega} 
V \eta V \frac{\lambda^1}{\omega} 
V \frac{\lambda^2}{\omega^2} V 
\frac{\lambda^1}{\omega} V \eta\nn
&+&\frac{1}{2}\eta 
A_\pi \frac{\lambda^1}{\omega} V \eta 
V \frac{\lambda^1}{\omega} V 
\frac{\lambda^2}{\omega} V 
\frac{\lambda^1}{\omega^2} V \eta+\eta A_\pi 
\frac{\lambda^1}{\omega} V 
\frac{\lambda^2}{\omega^2} V 
\frac{\lambda^1}{\omega} V \eta V 
\frac{\lambda^1}{\omega} V \eta+\eta A_\pi 
\frac{\lambda^1}{\omega} V 
\frac{\lambda^2}{\omega} V 
\frac{\lambda^1}{\omega^2} V \eta V 
\frac{\lambda^1}{\omega} V \eta\nn
&+&\frac{1}{2}\eta 
A_\pi \frac{\lambda^1}{\omega} V 
\frac{\lambda^2}{\omega} V 
\frac{\lambda^1}{\omega} V \eta V 
\frac{\lambda^1}{\omega^2} V \eta-\eta A_\pi 
\frac{\lambda^1}{\omega} V 
\frac{\lambda^2}{\omega} V 
\frac{\lambda^1}{\omega} V 
\frac{\lambda^2}{\omega} V 
\frac{\lambda^1}{\omega} V \eta-\eta A_\pi 
\frac{\lambda^1}{\omega} V 
\frac{\lambda^2}{\omega} V 
\frac{\lambda^3}{\omega} V 
\frac{\lambda^2}{\omega} V 
\frac{\lambda^1}{\omega} V \eta\nn
&-&\frac{1}{2}\eta 
V \frac{\lambda^1}{\omega^3} V \eta 
V \frac{\lambda^1}{\omega} V \eta 
A_\pi \frac{\lambda^1}{\omega} V 
\eta-\frac{1}{2}\eta V \frac{\lambda^1}{\omega^2} 
V \eta A_\pi \frac{\lambda^1}{\omega^2} 
V \eta V \frac{\lambda^1}{\omega} 
V \eta-\frac{1}{4}\eta V 
\frac{\lambda^1}{\omega^2} V \eta A_\pi 
\frac{\lambda^1}{\omega} V \eta V 
\frac{\lambda^1}{\omega^2} V \eta\nn
&+&\frac{1}{2}\eta 
V \frac{\lambda^1}{\omega^2} V \eta 
A_\pi \frac{\lambda^1}{\omega} V 
\frac{\lambda^2}{\omega} V 
\frac{\lambda^1}{\omega} V \eta-\frac{3}{8}\eta 
V \frac{\lambda^1}{\omega^2} V \eta 
V \frac{\lambda^1}{\omega^2} V \eta 
A_\pi \frac{\lambda^1}{\omega} V 
\eta+\frac{1}{2}\eta V \frac{\lambda^1}{\omega^2} 
V \eta V \frac{\lambda^1}{\omega} 
A_\pi \frac{\lambda^2}{\omega} V 
\frac{\lambda^1}{\omega} V \eta\nn
&+&\frac{1}{2}\eta 
V \frac{\lambda^1}{\omega^2} V \eta 
V \frac{\lambda^1}{\omega} V 
\frac{\lambda^2}{\omega} A_\pi 
\frac{\lambda^1}{\omega} V \eta+\frac{1}{2}\eta 
V \frac{\lambda^1}{\omega^2} V 
\frac{\lambda^2}{\omega} V 
\frac{\lambda^1}{\omega} V \eta A_\pi 
\frac{\lambda^1}{\omega} V \eta-\frac{1}{2}\eta 
V \frac{\lambda^1}{\omega} A_\pi \eta 
V \frac{\lambda^1}{\omega^3} V \eta 
V \frac{\lambda^1}{\omega} V 
\eta \nn
&+&\frac{1}{2}\eta V 
\frac{\lambda^1}{\omega} A_\pi \eta V 
\frac{\lambda^1}{\omega^2} V 
\frac{\lambda^2}{\omega} V 
\frac{\lambda^1}{\omega} V \eta+\frac{1}{2}\eta 
V \frac{\lambda^1}{\omega} A_\pi \eta 
V \frac{\lambda^1}{\omega} V 
\frac{\lambda^2}{\omega^2} V 
\frac{\lambda^1}{\omega} V \eta+\eta V \frac{
\lambda^1}{\omega} A_\pi 
\frac{\lambda^2}{\omega^2} V 
\frac{\lambda^1}{\omega} V \eta V 
\frac{\lambda^1}{\omega} V \eta\nn
&+&\eta V \frac{
\lambda^1}{\omega} A_\pi \frac{\lambda^2}{\omega} 
V \frac{\lambda^1}{\omega^2} V \eta 
V \frac{\lambda^1}{\omega} V \eta-\eta 
V \frac{\lambda^1}{\omega} A_\pi 
\frac{\lambda^2}{\omega} V 
\frac{\lambda^1}{\omega} V 
\frac{\lambda^2}{\omega} V 
\frac{\lambda^1}{\omega} V \eta-\eta V \frac{
\lambda^1}{\omega} A_\pi \frac{\lambda^2}{\omega} 
V \frac{\lambda^3}{\omega} 
V \frac{\lambda^2}{\omega} V 
\frac{\lambda^1}{\omega} V \eta\\
&+&\eta V \frac{
\lambda^1}{\omega} V \eta V \frac{\lambda^1}{
\omega^2} A_\pi \frac{\lambda^2}{\omega} 
V \frac{\lambda^1}{\omega} V \eta-\eta 
V \frac{\lambda^1}{\omega} V 
\frac{\lambda^2}{\omega} A_\pi 
\frac{\lambda^1}{\omega} V 
\frac{\lambda^2}{\omega} V 
\frac{\lambda^1}{\omega} V \eta-\eta V \frac{
\lambda^1}{\omega} V \frac{\lambda^2}{\omega} 
A_\pi \frac{\lambda^3}{\omega} 
V \frac{\lambda^2}{\omega} V 
\frac{\lambda^1}{\omega} V \eta+ \;{\rm h.c.}\;.
\nonumber
\eeqa

\end{document}